\documentstyle[preprint,prl,aps]{revtex}
\begin{document}
\draft
\title{Tri-critical behavior in rupture induced by disorder}
\author{J\o rgen Vitting Andersen$^1$, Didier Sornette$^{1,2}$ and Kwan-tai
Leung$^3$}
\address{$^1$ Laboratoire de Physique de la Mati\`{e}re Condens\'{e}e, CNRS
URA 190\\
Universit\'{e} de Nice-Sophia Antipolis, Parc Valrose, 06108 Nice, France}
\address{$^2$ Department of Earth and Space Sciences and Institute of
Geophysics and Planetary Physics\\ University of California, Los Angeles,
California
90095-1567}
\address{$^3$ Institute of Physics, Academia Sinica, Nankang, Taipei,
Taiwan 11529,
R.O.C. }

\date{\today}
\maketitle
\begin{abstract}
We discover a qualitatively new behavior for systems where the
load transfer has limiting stress amplification as in real fiber
composites. We find that
the disorder is a relevant field leading to tri--criticality, separating a
first-order regime where rupture occurs without significant precursors from
a
second-order regime where the macroscopic elastic coefficient
exhibit power law behavior. Our results are based on analytical
analysis of
fiber bundle models and numerical simulations of a two-dimensional
tensorial
spring-block system in which stick-slip motion and fracture compete.
\end{abstract}


There is growing evidence that rupture in random media can be viewed
as a kind of critical phenomenon \cite{Herrmann,Vanneste} as a result of
the interplay
between disorder and fracture mechanics, with proposed applications in
particular to
fiber composites \cite{Anifrani,Zhou} and earthquakes \cite{critearth}.
Notwithstanding its importance, we do not have a comprehensive
understanding of
rupture phenomena but only a partial  classification
\cite{Herrmann,Vanneste,Roux}.
>From a theoretical point of view, rupture is controlled in principle by the
infinite
moment $\langle \sigma^q \rangle|_{q \to \infty}$ of the stress field and
the
difficulties emerge from the non-commutation of the two limits $(q \to
\infty, \Delta
\to 0)$, where $\Delta$ is the amount of disorder (see below for a precise
definition).  In intuitive wording, the largest stress in the system is
very sensitive
to the amount and type of disorder. Disorder is known to induce stress
field
distributions with fat tails \cite{distrib}. Consider for instance a
log-normal distribution with
standard deviation $\Delta$ and mean $\sigma_0$, then
$\langle \sigma^q \rangle^{1/q} = \sigma_0 e^{\Delta^2 q/2}$,
which shows that the limit
$\Delta \to 0$
is singular for rupture ($q \to \infty$). This non-commutativity of limits
is at the
crux of some of the major outstanding problems in physics \cite{Berry} such
as turbulence (viscosity $\to 0$ ; time $\to \infty$) and quantum chaos ($h \to
0$ ; time $\to \infty$). Here, we show that the amount of disorder $\Delta$
plays the role of a relevant field which makes systems with limited stress
amplification exhibit a tri-critical transition as the disorder increases, from
a Griffith-type abrupt rupture (first-order) regime to a progressive damage
(critical) regime. This is reminiscent of the critical behavior induced by
quenched disorder in magnetic systems \cite{Berker}.

We first document this behavior in a simple mean-field model of rupture,
known as the
democratic fiber bundle model \cite{DFBM}.
It consists of $N$ parallel fibers with identical spring
constants and identically  independent random failure thresholds
$X_j$
distributed
according to the cumulative probability distribution $P(X_j < x) \equiv
P(x)$. A total
force $F$ is applied to the system and is shared democratically among the
$N$ fibers.
When the force on one fiber reaches its threshold, the fiber ruptures and
the stress is
redistributed to all remaining fibers. This transfer might induce
secondary failures which in turn induce tertiary ruptures and so on. One is
interested in the stress-strain characteristic as the applied force is
increased, the
properties of the rupture point and the precursory events prior to the
complete
breakdown. The solution of this problem is found by noticing that the total
bundle
will not break under a load $F$ if there are $n$ fibers in the bundle each
of
which can withstand $x_n \equiv F/n$. $x_n$ and $n$ are related,
for large $N$, by $n = N [1-P(x_n)]$
leading to $F(x_n) = N x_n [1- P(x_n)]$. The number $k$ of fibers which have
failed under the
force $F$ is then $k = N-n = N P(x_n)$. Now, for a broad class of
distribution $P(x)$
extending down to $0$, the function $x [1- P(x)]$ presents a
maximum at $0 < x^* < \infty$,
the solution of $d x [1- P(x)]/dx|_{x=x^*} = 0$.
As the behavior of $F(x_n)$ close to $x^*$ is quadratic
$F(x_n) \approx F^* - c (x^* - x_n)^2$
where $c$ is a constant, this implies that the rate $dk/dF$ of
fiber failure
diverges as $(F^* - F)^{-1/2}$, where $F^* = x^* [1- P(x^*)]$, thus
qualifying a critical mean field behavior.

However, if $P(x)$ is such that $d x [1- P(x)] /dx = 0$ has no
solution, the behavior will be completely different, with a sequence of a
few fibers
maybe breaking as the load is applied followed by an abrupt global failure.
Correspondingly, the stress-strain characteristic exhibits a discontinuity
in its
slope at the point of rupture. This qualifies a first-order behavior.
Notice that this
is similar to the Ehrenfest's classification of the order of phase
transitions, where
the free energy is here replaced by the elastic energy.

Let us take for
instance $P(x) = 0$ for $0 \leq x < x_1$,
$P(x) = (x-x_1)/\Delta$ for $x_1 \leq x \leq x_1 + \Delta$ and $P(x)
= 1$ for
$x \geq x_1 + \Delta$, corresponding to the strengths $X_j$ uniformly
distributed
between $x_1 > 0$ and $x_1 + \Delta$.
Then,  $d x [1- P(x)] /dx =
(x_1 + \Delta - 2 x)/ \Delta$,
which has a root in the interval $x_1 \leq x \leq x_1 + \Delta$
if and only
if $\Delta > x_1$. In this case, we recover the previous mean field
critical behavior.
However, for  weak disorder $\Delta < x_1$, not a single fiber breaks down
until
the force reaches $N x_1$ at which value the system of $N$ fibers breaks
suddenly.
This is an extreme illustration of a ``first-order'' behavior. The
particular value
$\Delta = x_1, F=Nx_1$ thus plays the role of a tri-critical point in
analogy with
thermal phase transitions \cite{tricri}. This behavior holds for a large
class of
distributions $P(x)$: the condition that $d x [1- P(x)] /dx$ has no
root is
equivalent to the condition that the equation
$d\log [1- P(x)] /dx = -1/x$
has no solutions for any $0 \leq x < \infty$. This equation defines two
domains:
(1)
$d\log [1- P(x)] /dx < -1/x$
for all $x \geq 0$: this can occur in
particular if $1- P(x)$ decays to zero faster than $1 /x$ for large
$x$, with
the additional constraint that there exists a minimum strength $x_1$
strictly
positive. Notice that the distributions which extend down to zero are in
this
sense always in the ``large'' disorder regime.
(2)
$d\log [1- P(x)] /dx > -1/x$
for all $x \geq 0$: this corresponds to distributions which
decays slower
than $1 /x$. Take for instance $1-P(x) = (1+x)^{-\alpha}$.
Then,
$d\log [1- P(x)] /dx = -\alpha /(1+x)$ which remains strictly
larger than
$-1/x$ if  $\alpha < 1$.

Having established the existence of the tri-critical behavior in this mean
field model,
let us now turn our attention to a more realistic two-dimensional
(2D) spring-block model of
surface
fracture in which the stress can be released by spring breaks {\it and\/}
block slips.
We consider the
experimental
situation where a balloon covered with paint or dry resin is progressively
inflated\cite{note2}.
An industrial application is a metallic tank with carbon or kevlar fibers
impregnated
in a resin matrix wrapped up around it  which is slowly
pressurized \cite{Anifrani}. As
a consequence, it elastically deforms, transferring tensile stress to the
overlayer.
Slipping (called fiber-metal delamination) and cracking can thus occur in
the
overlayer. We model this process by a 2D array of blocks which represents
the
overlayer on a coarse grained scale in contact with a surface with solid
friction contact.
The solid friction will limit stress amplification.
Each block is interconnected to its nearest neighbors via springs
of
unstretched lengths $l_0$ and spring constants $K$. The position of the
blocks in the
$x$- and $y$-directions are  given by $(a \cdot i+x_{i,j}, a \cdot
j+y_{i,j})$ where
$1\leq i,\,j \leq L$,  form a square lattice with lattice constants $a$,
and where
$x_{i,j}$, $y_{i,j}$ fulfill $x_{i,j}, y_{i,j} \ll a$,  so that Hooke's law
applies.
In \cite{Tai} it was shown that the $x$ component of the force on a block
{\em to
first order}
in the displacements takes the form:
\begin{eqnarray}
F^x_{i,j}&=& - K \{ (b_{i+1,j}+b_{i-1,j})x_{i,j} - b_{i+1,j}x_{i+1,j} -
\nonumber \\ & &
b_{i-1,j}x_{i-1,j} +  s[(b_{i,j+1}+b_{i,j-1})x_{i,j} -
b_{i,j-1}x_{i,j-1}
\nonumber \\
& & - b_{i,j+1}x_{i,j+1} ]  -
as(b_{i+1,j} - b_{i-1,j})\}
\label{forcexcomponent}
\end{eqnarray}
and, by symmetry, $F^y_{i,j}$ follows by switching
$x \leftrightarrow y$ and $i \leftrightarrow j$.
$s \equiv (a - l_0)/ a\geq 0$
is the strain of the network without fluctuations ($x_{i,j},y_{i,j}\equiv 0$)
, and $b_{i\pm 1,j\pm 1}=1, 0$,
respectively,
depending on whether a spring connects the blocks $(i,j)-(i\pm 1,j \pm 1)$
or not.
Likewise the stress $B$ in a spring is given by:
\begin{eqnarray}
B_{(i,j)-(i\pm 1,j)}&=&
 K [ (x_{i,j}-x_{i\pm 1,j}-s)^2 +
\nonumber \\ & &
s^2 (y_{i,j}-y_{i\pm 1,j})^2 ]^{1/2}
\label{bondstress}
\end{eqnarray}
Initially $x_{i,j}$ and $y_{i,j}$ are chosen uniformly from the interval
$[-\Delta, +\Delta]$, thus $\Delta$ quantifies the amount of disorder which is
on the initial displacements corresponding to an effective initial
disorder in the thresholds. Periodic
boundary conditions are used in both the $x$-- and $y$--direction.

The coupling of the overlayer to the substrate has two effects
when the substrate expands:
(1)
Tensile stress is transferred to the overlayer. This is taken into account
by imposing an increase in the average
distance
$a$ between the blocks so as to reflect the inflation of the balloon.
As a definition of the time, $t$, we let $a(t)$ increase linearly with $t$.
(2) The increasing tensile stress
 in turn gives rise to stick-and-slip motion or/and cracking.
A block is  assumed to stick until the total force applied on
it exceeds a threshold $F_s$, where after it slips to the zero-force
position,
corresponding to local mechanical equilibrium in the absence of friction. This
thereby
releases stresses on its neighbor blocks. A spring breaks irreversibly once
the stress $B$ exceeds a threshold $F_c\equiv \kappa F_s$\cite{note1}.

We define a time dependent apparent macroscopic stress on the system,
$\sigma_{app} (t)$,
from the relation  $\sigma_{app} (t) = {E(t) /\epsilon (t)}$, where
$E(t)$ is the
total elastic energy stored in the springs of the system at time $t$, and
$\epsilon
(t) \equiv a(t)/a(t=0)$ is the macroscopic strain.
We calculate an effective Young modulus, given by
\begin{equation}
Y_{app}(t) \equiv {d\sigma_{app}(t) /d\epsilon}.
\label{young}
\end{equation}
$Y_{app}(t)$ can be expressed as
$[d\sigma_{app}(t) / d\sigma(t)] (d\sigma /d\epsilon)$,
where $\sigma$ is proportional to
the
first invariant (the trace) of the real stress field in the system.
${d\sigma /d\epsilon}$ is the corresponding elastic modulus
expected to
exhibit a power law behavior if criticality is present, while
${d\sigma_{app}(t) /d\sigma(t)}$
goes to a constant. Therefore, the measurement of $Y_{app}(t)$
gives us
direct access, if present, to the critical behavior of the Young modulus of
the system.
As the strain $\epsilon(t)$ is increased, block slips and spring failures
occur up
to a point
where the system is completely ruptured and the stress necessary to impose
a constant
small strain rate starts to decrease from a maximum. At this point, there
is at least
one large crack spanning the whole system. If global rupture occurs abruptly
(first-order
case),  $\sigma_{app} (t)$ must exhibit a sharp maximum and  $Y_{app}(t)$
remains
finite. If, on the other hand, the rupture is critical,  $\sigma_{app} (t)$
will
exhibit a progressive rounding with a smooth maximum, while $Y_{app}(t)$
vanishes as a
power law $Y_{app} \propto[(\epsilon_c - \epsilon)/\epsilon_c]^{\gamma}$ on
the approach to rupture at $\epsilon_c$.

In Fig.~1 are shown the stress--strain curves
from one single realization for each different system size,
with $\Delta=0.75\Delta_c$, where $\Delta_c$ is the
maximal amplitude such that at $t=0$ $B_{(i,j)-(i\pm 1,j\pm 1)} < F_c$ and
$| \vec{F}_{i,j} |< F_s$ for all $(i,j)$. $\Delta_c$ has been determined
numerically for each run.
We set $a(t=0)=lo=1$ and $Fs/K=1$ throughout this paper.
Observe that the maximum stress a
system can sustain is an increasing function of $\kappa$ and the
range of $\epsilon$-values over which fracturing takes place decreases
as $\kappa$ increases. For $\Delta=0.75\Delta_c$ and $\kappa < 2.9$, the
stress--strain curve presents a smooth maximum indicating a critical
rupture. This is
confirmed in Fig.~2,
showing the vanishing of the apparent Young modulus
$Y_{app}$ as $\epsilon\to\epsilon_c$ from below.
Each curve is obtained
by averaging over $N=1000-5000$ independent configurations with system size
$L=30$.
$\epsilon_c$ has been estimated from the condition
$d \langle \sigma_{app} \rangle /d \epsilon|_{\epsilon  =\epsilon_c} =0$,
where $\langle \cdots \rangle$
stands for an ensemble average. Larger lattice sizes $(L=50-400)$ with the same
value of $LN$ were used, but for a given fixed $LN$ we found the  smallest
lattice sizes $L$ ($=30$) give the best statistics,  which we attribute to the
lack of self averaging.  For small $\kappa$, the exponent $\gamma$ approaches a
value slightly  larger than $1$, while it decreases continuously to zero as
$\kappa$ increases, as shown in the inset. It seems to vanish around $\kappa =
2.9$, signaling the transformation of the critical regime to an abrupt
``first-order'' behavior. Keeping $\kappa$ fixed and varying ${\Delta
/\Delta_c}$, we find
that $\gamma$ stays constant, but the size of the critical region increases with
the magnitude of $\Delta/\Delta_c$. It shrinks to zero at a threshold value
function of
$\kappa$ which is shown in Fig.~3.  This function gives the
boundary in the $(\Delta /\Delta_c; \kappa)$  plane between the critical
and first-order
regime. As announced, for fixed $\kappa < 2.9$, increasing the disorder $\Delta$
allows the system to go from a first-order to a critical regime. The fact that
the disorder is so relevant as to create the analog of a tri-critical behavior
can be tracked back to the existence of solid friction on the blocks which
ensures that the elastic forces in the springs are carried over a bounded
distance (equal to the size of a slipping ``avalanche'') during the stress
transfer induced by block motions.

When $\kappa$ is large, the system responds initially to an expansion by
the release of stress uniquely by block slips. The block slips give rise
to a stress rearrangement, and a spatial coarsening phenomenon of the
stress field $B$ takes place \cite{Joergen1}. The enhanced
correlations in the $B$-field result in turn in a coherence
when fracturing sets in, amounting to smoothing out the disorder,
thus allowing for a large crack to develop in an abrupt way. For sufficiently
large
$\kappa$, the system breaks into two parts. This is the regime of first-order
behavior. Notice that increasing $\kappa$ in the $(\Delta /\Delta_c; \kappa)$
phase diagram corresponds to decreasing the disorder and changing at the same
time the distribution of disorder so that it becomes more
correlated. This is therefore a more complicated route than just changing the
width of the threshold distribution as in the previous fiber bundle model.

The value of the Young critical exponent $\gamma$ for small $\kappa$ can be
predicted from percolation theory.  Indeed, consider the limit $\kappa \to 0$,
for which the blocks are stuck to the substrate and cannot move. Only the
springs
can fail and they do so in a completely uncorrelated way, controlled by the
initial random configuration of the blocks. We thus get an uncorrelated random
dilution, ending at the percolation threshold where a macroscopic crack spans
the system. In the presence of internal strain, it was shown that the elastic
constant decreases to zero when the dilution increases with an exponent given by
the {\it scalar} elasticity problem \cite{Alexander}, equal to the conductance
exponent of percolation. Extensive numerical simulations give the value $1.300$
for this exponent \cite{Herr}, which is in agreement with the extrapolation
of our results for $\kappa \to 0$.

It is important to understand that these properties belong to systems with load
transfer mechanisms limiting stress amplification at crack tips.
If no coupling or delay mechanism exist to reguralize the divergence
induced by elasticity at the crack tips (with a stress diverging as
$1/\sqrt{r}$ next to a crack tip), the first-order
behavior is only observed for zero disorder described by
the single-crack Griffith criterion and any amount of
disorder is relevant to produce a critical behavior \cite{Herrmann,Vanneste}.
However, even in this case, the amount of
disorder remains of utmost importance as it controls the size of the critical
region, and therefore its observability \cite{VVSS}.

The existence of different regimes for rupture, depending on the limiting
stress amplification and on disorder, opens the road to important potential
applications for failure prediction purposes such as in the time-to-failure
approach \cite{timeto}. We suggest that the often observed power law
distribution of acoustic emission bursts of many materials upon stressing,
offers an additional evidence of the critical nature of the  damage and
cracking of heterogeneous materials \cite{Pollock,Anifrani}.
Our results provide the foundation for understanding why
some systems exhibit clearer precursors before rupture than others in which
they may even be absent in certain cases and for
quantifying the expected amount and style of precursory activity as a function
of heterogeneity and range of interaction.

\vspace{0.5in}
J.V.A wishes to acknowledge support from
the Danish Natural Science Research Council under Grant No. 9400320, and
support from the European Union Human Capital and Mobility Program
contract number ERBCHBGCT920041 under the direction of Prof. E. Aifantis.
K.-t. L. is supported by the National Science Council of ROC.


\begin{figure}
\begin{center}
\caption{
Stress-strain curves for different values of
$\kappa$ and for
different system sizes $L = 50$ (dotted line),
$100$ (thin bold line), and $200$ (bold line).
The fracturing is stopped at different $\epsilon (t)$ for
different $L$ in order to distinguish between the curves.
One $L=400$ simulation has been done for $\kappa = 4$ (fat bold line).
}
\end{center}
\label{stress_strain}
\end{figure}
\begin{figure}
\begin{center}
\caption{
Macroscopic Young modulus
vs reduced macroscopic strain
for different values of
$\kappa =0.5 (\diamond)$, $0.75 (+)$, $0.875 (\Box)$, $1.4 (\times)$  and
$2.0 (\triangle)$.
$\Delta /\Delta_c=0.75$.
The inset shows the exponent
$\gamma$ as a function of $\kappa$. $\kappa \approx 0.5$
is the smallest value for which the system initially has no bonds
that exceed the threshold.
}
\end{center}
\label{power_law}
\end{figure}
\setcounter{figure}{1}
\begin{figure}
\begin{center}
\caption{
Inset to Fig.~2
}
\end{center}
\label{gamma}
\end{figure}
\begin{figure}
\begin{center}
\caption{
Phase diagram for criticality of the
fracturing. $\Delta/\Delta_c \to 0$ can not been studied within the
spring--block model
since ambiguity in the updating rules for stress release,
affects the fracturing.
}
\end{center}
\label{phase_diagram}
\end{figure}


\begin{references}
\bibitem{note2}
The case where the stiffness of an overlayer changes on a fixed rough
substrate has been studied in \cite{Joergen1}.
\end{references}
\end{document}